\documentclass[pra,twocolumn,superscriptaddress,10pt]{revtex4-1}
\usepackage{graphicx}
\usepackage{dcolumn}
\usepackage{color}
\usepackage{times}
\usepackage{bm}
\usepackage{amssymb}
\usepackage{amsmath}
\usepackage{epsfig}
\usepackage{epstopdf}
\usepackage{dsfont}
\usepackage{float}
\usepackage{multirow}
\usepackage[colorlinks=true,citecolor=blue, linkcolor=blue]{hyperref}

\begin{document}
	
	\title{\textbf{Experimental investigation of entropic uncertainty relations and coherence uncertainty relations}}

	\author{Zhi-Yong Ding}
	\affiliation{School of Physics and Material Science, Anhui University, Hefei 230601, China}
	\affiliation{School of Physics and Electronic Engineering, Fuyang Normal University, Fuyang 236037, China}
	\affiliation{Key Laboratory of Functional Materials and Devices for Informatics of Anhui Educational Institutions, Fuyang Normal University, Fuyang 236037, China}
		
	\author{Huan Yang}
	\affiliation{School of Physics and Material Science, Anhui University, Hefei 230601, China}
	\affiliation{Institutes of Physical Science and Information Technology, Anhui University, Hefei 230601, China}
	\affiliation{Department of Experiment and Practical Training Management, West Anhui University, Lu'an 237012, China}
		
	\author{Dong Wang}
	\email[]{dwang@ahu.edu.cn (D. Wang)}
	\affiliation{School of Physics and Material Science, Anhui University, Hefei 230601, China}
	\affiliation{CAS Key Laboratory of Quantum Information, University of Science and Technology of China, Hefei 230026, China}
	
	\author{Hao Yuan}	
	\affiliation{School of Physics and Material Science, Anhui University, Hefei 230601, China}
	\affiliation{CAS Key Laboratory of Quantum Information, University of Science and Technology of China, Hefei 230026, China}
	\affiliation{Key Laboratory of Opto-Electronic Information Acquisition and Manipulation of Ministry of Education, Anhui University, Hefei 230601, China}
		
	\author{Jie Yang}
	\affiliation{School of Physics and Material Science, Anhui University, Hefei 230601, China}
	
	\author{Liu Ye}
	\email[]{yeliu@ahu.edu.cn (L. Ye)}
	\affiliation{School of Physics and Material Science, Anhui University, Hefei 230601, China}

\date{\today}
	
	\begin{abstract}
		 Uncertainty relation usually is one of the most important features in quantum mechanics, and
is the backbone of quantum theory, which distinguishes from the rule in classical counterpart. Specifically, entropy-based uncertainty relations are
 of fundamental importance in the region of quantum information theory,
offering one nontrivial bound of key rate towards quantum key distribution. In this work, we
experimentally demonstrate the entropic uncertainty relations and coherence-based uncertainty relations in an all-optics platform.
By means of preparing two kinds of bipartite initial
states with high fidelity, i.e., Bell-like states  and Bell-like diagonal states, we carry on local projective measurements over
a complete set of mutually unbiased bases  on the measured subsystem. In terms of quantum tomography, the density matrices of the initial states
and the post-measurement states are reconstructed. It shows that our experimental results
coincide with the theoretical predictions very well. Additionally, we also verify that the lower bounds of both the entropy-based and
coherence-based uncertainty can be tightened by imposing the Holevo quantity and mutual information, and the entropic uncertainty is inversely
correlated with the coherence. Our demonstrations might offer an insight into their uncertainty relations and their connection to quantum coherence in quantum information science,
which might be applicable to the security analysis of quantum key distributions.
	\end{abstract}

	\maketitle
	
	\section{INTRODUCTION}
Heisenberg's uncertainty principle \cite{Heisenberg172} is deemed as one of the most fundamental features of  quantum world,
which essentially is different from
classical world. The uncertainty principle manifests that it is impossible to accurately predict both the position and the momentum of a microscopic particle.
In other words, the more accurate the position measurement is, the more uncertain the momentum measurement will be, and vice versa \cite{Coles015002, Wang1900124}.
Before long, Robertson \cite{Robertson163} generalized the uncertainty principle to an arbitrary pair of non-commuting observables. Notably,
 there is a conceptual shortcoming in Robertson's inequality, leading to a trivial result when the systemic state
 is in the state of one of eigenstates of the measured observables.
Afterwards, Deutsch had established the
celebrated entropic uncertainty relation (EUR) \cite {Deutsch631, Kraus3070}, in terms of Shannon entropy, which was improved by Maassen and Uffink into the form of \cite{Maassen1103}
		\begin{equation}\label{g01}
			H(P) + H(Q) \ge {\log _2}\frac{1}{c},
		\end{equation}
		where $P$ and $Q$ are two arbitrary observables, $H( \cdot )$  denotes the Shannon entropy, and $c = {\max _{i,j}}{\left| {\left\langle {{p_i}|{q_j}} \right\rangle } \right|^2}$
stands for the maximum overlap between any two eigenvectors ($\{ \left| {{p_i}} \right\rangle \} $ and $\{ \left| {{q_j}} \right\rangle \} $) of $P$ and $Q$.

Notably, the entropy-based uncertainty relation mentioned previously is suitable for observing the  law of incompatible measurements in a single-particle system.
As to a composite system ${\hat\rho}_{AB}$ with $A$ to be probed and  $B$ as a quantum memory, Berta \textit{et al}. \cite{Berta659}
had put forward   the quantum-memory-assisted entropic uncertainty relation, which is given by
		\begin{equation}\label{g02}
			S(P|B) + S(Q|B) \ge {\log _2}\frac{1}{c} + S(A|B),
		\end{equation}
		where $S(A|B) = S({{\hat\rho} _{AB}}) - S({{\hat\rho} _B})$ represents the conditional von Neumann entropy of ${\hat\rho}_{AB}$ and ${{\hat\rho} _B} = {{\rm{Tr}}_A}{{\hat\rho} _{AB}}$ is the reduced state, $S(P|B) = S({{\hat\rho} _{PB}}) - S({{\hat\rho} _B})$  denotes the conditional von Neumann entropy of the post-measurement state ${\hat\rho}_{PB}$,	
which can be obtained by performing the local projective measurements $\{ \left| {{p_i}} \right\rangle \left\langle {{p_i}} \right|\} $ on the subsystem $A$.
Subsequently, Li \textit{et al.} \cite{Li752} and Prevedel \textit{et al.} \cite{Prevedel757} experimentally demonstrated the quantum-memory-assisted entropic uncertainty relation
through two back-to-back all-optical configurations.

Technically, EURs had yielded many potential applications in the domain of quantum information science, e.g.,
quantum key distribution \cite{Berta659,Li752}, entanglement witness \cite{Berta659,Li752,Prevedel757,Coles022112}, quantum teleportation \cite{Hu032338}, quantum steering \cite{Schneeloch062103,Walborn130402}, and quantum metrology \cite{Jarzyna013010}. Accordingly, EURs has received much attention
and several tighter lower bounds  of  the uncertainty were proposed by means of various methods \cite{Pati042105, Hu014105, Coles022112, Adabi062123, Huang545}.
Later on, the entropic uncertainty relations
are generalized to multi-observable versions \cite{Liu042133, Xiao042125}. Soon afterwards, Xing \textit{et al.} \cite{Xing2563} reported an experimental investigation of entropic
uncertainty relations with regard to three measurements in a pure diamond system. In addition, there exhibit some promising experiments to verify different types
of  uncertainty relations \cite{Ma160405, Xiao17904, Wang39, Chen062123}. 		
		
		On the other hand, quantum coherence, the embodiment of the superposition principle of quantum states, is also one of the most fundamental features that marks the violation of quantum mechanics from the classical world \cite{Streltsov041003, Hu1, Vicente045301, Marvian052324}. It has been widely concerned and studied since Baumgratz \textit{et al.} \cite{Baumgratz140401} established a rigorous framework for the quantification of coherence. Generally, quantum coherence is related to the characteristics of the whole system. In order to reveal the relationship between quantum coherence and quantum correlations, the unilateral coherence \cite{Ma160407, Hu052106} is introduced for a bipartite system. Intrinsically, quantum coherence of a quantum state depends on the choice of the reference basis. If two or more incompatible reference bases are selected, coherence-based uncertainty relations (CURs) can be established \cite{Singh47, Yuan032313, Dolatkhah13, Fan085203}. Soon after, Lv \textit{et al.} \cite{Lv062337} demonstrated an all-optical experiment of CURs in two different reference bases.
				
		In this paper, we focus on achieving experimental investigations with respect to EURs and  CURs in
an all-optics framework. By means of preparing two kinds of bipartite initial states, i.e., Bell-like states (pure) and Bell-like diagonal states (mixed),
 we then perform local projective measurements over a complete set of mutually unbiased bases (MUBs) on one of the subsystem \cite{Wootters391, Wootters363}.
 By utilizing quantum tomography \cite{James052312}, we obtain the density matrices of the initial states and the post-measurement states,
 as well as the corresponding measurement probability. It shows our experimental results coincide with the theoretical predictions very well.
Moreover, we also demonstrate that the lower bounds of both the entropic and coherence-based uncertainty can be strengthened based on the Holevo quantity and mutual information.
	
	\section{THEORETICAL FRAMEWORK}
		Briefly, we herein review the EURs for multiple measurements, and then render a theoretical framework of the CURs. For a quantum state ${\hat\rho} $ in a $d$-dimensional Hilbert space ${\cal H }$, once resorting to an arbitrary set of basis $\{ \left| i \right\rangle \} $, the relative entropy of coherence can be written as \cite{Baumgratz140401}
		\begin{equation}\label{g03}
			C_{{\rm{re}}}^{(i)}({\hat\rho} ) = S({\hat{\rho}} _{{\rm{diag}}}^{(i)}) - S({\hat{\rho}} ),
		\end{equation}
		where $S({\hat\rho} ) =  - {\rm{Tr}}({\hat\rho} \log {\hat\rho} )$ is the von Neumann entropy, and ${\hat\rho} _{{\rm{diag}}}^{(i)} = \sum\nolimits_i {\left\langle i \right|{\hat\rho} \left| i \right\rangle \left| i \right\rangle \left\langle i \right|} $ is the diagonal part of ${\hat\rho} $ in the basis $\{ \left| i \right\rangle \} $. For a bipartite state ${{\hat\rho} _{AB}}$ in a composite Hilbert space ${{\cal H}_A} \otimes {{\cal H}_B}$, after performing a local projective measurement $M = \{ \left| {{u_i}} \right\rangle \left\langle {{u_i}} \right|\} $ on subsystem $A$, the probability of obtaining result $i$   is quantified by ${p_i} = {\rm{Tr}}[({\left| {{u_i}} \right\rangle _A}\left\langle {{u_i}} \right| \otimes {\mathbb{I} _B}){{\hat\rho} _{AB}}]$ and the post-measurement state of the bipartite composite system reads as ${\hat\rho} _{AB}^i = ({\left| {{u_i}} \right\rangle _A}\left\langle {{u_i}} \right| \otimes {\mathbb{I} _B}){{\hat\rho} _{AB}}({\left| {{u_i}} \right\rangle _A}\left\langle {{u_i}} \right| \otimes {\mathbb{I} _B})/{p_i}$, where ${\mathbb{I} _B}$ is the identity operator in Hilbert space ${{\cal H}_B}$ of subsystem $B$.
As a result, the post-measurement state of the system is a mixture of $\{ {p_i},{{\hat\rho} _{AB}^i}\} $, explicitly given by
		\begin{equation}\label{g04}
			{{\hat\rho} _{MB}} = \sum\nolimits_i {({{\left| {{u_i}} \right\rangle }_A}\left\langle {{u_i}} \right| \otimes {\mathbb{I}_B}){{\hat\rho} _{AB}}({{\left| {{u_i}} \right\rangle }_A}\left\langle {{u_i}} \right| \otimes {\mathbb{I}_B})}.
		\end{equation}
		
In general, the unilateral coherence of the bipartite system can be written as \cite{Ma160407,Hu052106}
		\begin{equation}\label{g05}
			C_{{\rm{re}}}^M({{\hat\rho} _{AB}}) = S({{\hat\rho} _{MB}}) - S({{\hat\rho} _{AB}}).
		\end{equation}
		
		Furthermore, for a bipartite state ${{\hat\rho} _{AB}}$ and $N$ projective measurements ${M_m} = \{ \left| {u_{{i_m}}^m} \right\rangle \left\langle {u_{{i_m}}^m} \right|\} \;(m = 1, \cdots ,N)$ performed on subsystem $A$, the entropic uncertainty relations with regard to multiple measurements can be described by \cite{Liu042133}
		\begin{equation}\label{g06}
			\sum\limits_{m = 1}^N {S({M_m}|B)}  \ge {\log _2}\frac{1}{b} + (N - 1)S(A|B),
		\end{equation}
		 %$S(A|B)$ and $S({M_m}|B)$ are the conditional entropies of the initial state ${{\hat\rho} _{AB}}$ and the post-measurement state ${{\hat\rho} _{{M_m}B}}$ respectively,
here $ b = \mathop {\max }\limits_{{i_N}} \left\{ {\sum\nolimits_{i_2}^{i_{N - 1}} {\mathop {\max }\limits_{{i_1}} [ {c( {u_{{i_1}}^1,u_{{i_2}}^2} )} ]\prod\nolimits_{m = 2}^{N - 1} {c( {u_{{i_m}}^m,u_{{i_{m + 1}}}^{m + 1}} )} } } \right\} $ with  $c = \mathop {\max }\limits_{{i_1},{i_2}} c(u_{{i_1}}^1,u_{{i_2}}^2) = \mathop {\max }\limits_{{i_1},{i_2}} |\left\langle {u_{{i_1}}^1} \right|\left. {u_{{i_2}}^2} \right\rangle {|^2}$ denoting the largest overlap between any two eigenvectors of the given observables. According to Eqs. \eqref{g05} and \eqref{g06}, we can obtain the uncertainty relations for unilateral coherence under multiple measurements \cite{Dolatkhah13}, {\it viz.}
		\begin{equation}\label{g07}
			\sum\limits_{m = 1}^N {C_{{\rm{re}}}^{{M_m}}({{\hat\rho} _{AB}})}  \ge {\log _2}\frac{1}{b} - S(A|B),
		\end{equation}
		
		Specifically, for a $d \times d$-dimensional bipartite state ${{\hat\rho} _{AB}}$, if we select a complete set of ${d + 1}$  MUBs applied on subsystem $A$,
the EURs and CURs in MUBs can be derived into
		\begin{subequations}
			\begin{align}
				&\sum\limits_{m = 1}^{d + 1} {S({M_m}|B)}  \ge {\log _2}d + d\cdot S(A|B),\label{g08a} \\
				&\sum\limits_{m = 1}^{d + 1} {C_{{\rm{re}}}^{{M_m}}({{\hat\rho} _{AB}})}  \ge {\log _2}d - S(A|B),\label{g08b}
			\end{align}
		\end{subequations}
respectively. Note that,  the lower bound of the EURs can be strengthened by adding a term relating to the Holevo quantity and mutual information \cite{Dolatkhah13, Adabi062123}
		\begin{equation}\label{g09}
			\sum\limits_{m = 1}^{d + 1} {S({M_m}|B)}  \ge {\log _2}d + d\cdot S(A|B) + {\rm{Max}}\{ 0,\delta \},
		\end{equation}
		where $\delta  = (d + 1)I(A:B) - \sum\nolimits_{m = 1}^{d + 1} {I({M_m}:B)} $, $I(A:B) = S({{\hat\rho} _A}) + S({{\hat\rho} _B}) - S({{\hat\rho} _{AB}})$ is the mutual information of the bipartite system ${{\hat\rho} _{AB}}$, and $I({M_m}:B) = S({{\hat\rho} _B}) - \sum\nolimits_{{i_m}} {{p_{{i_m}}}S({{\hat\rho} _{{B_m}}})} $ is the Holevo quantity \cite{Nielsen}, which is an upper bound on the accessible information by the local projective measurement on the subsystem $A$.
Likewise, this additional term has been verified to be available on improving the lower bound of the CURs \cite{Fan085203}
		\begin{equation}\label{g10}
			\sum\limits_{m = 1}^{d + 1} {C_{{\rm{re}}}^{{M_m}}({{\hat\rho} _{AB}})}  \ge {\log _2}d - S(A|B) + {\rm{Max}}\{ 0,\delta \}.
		\end{equation}

\begin{figure*}
			\centering
			\includegraphics[width=12cm]{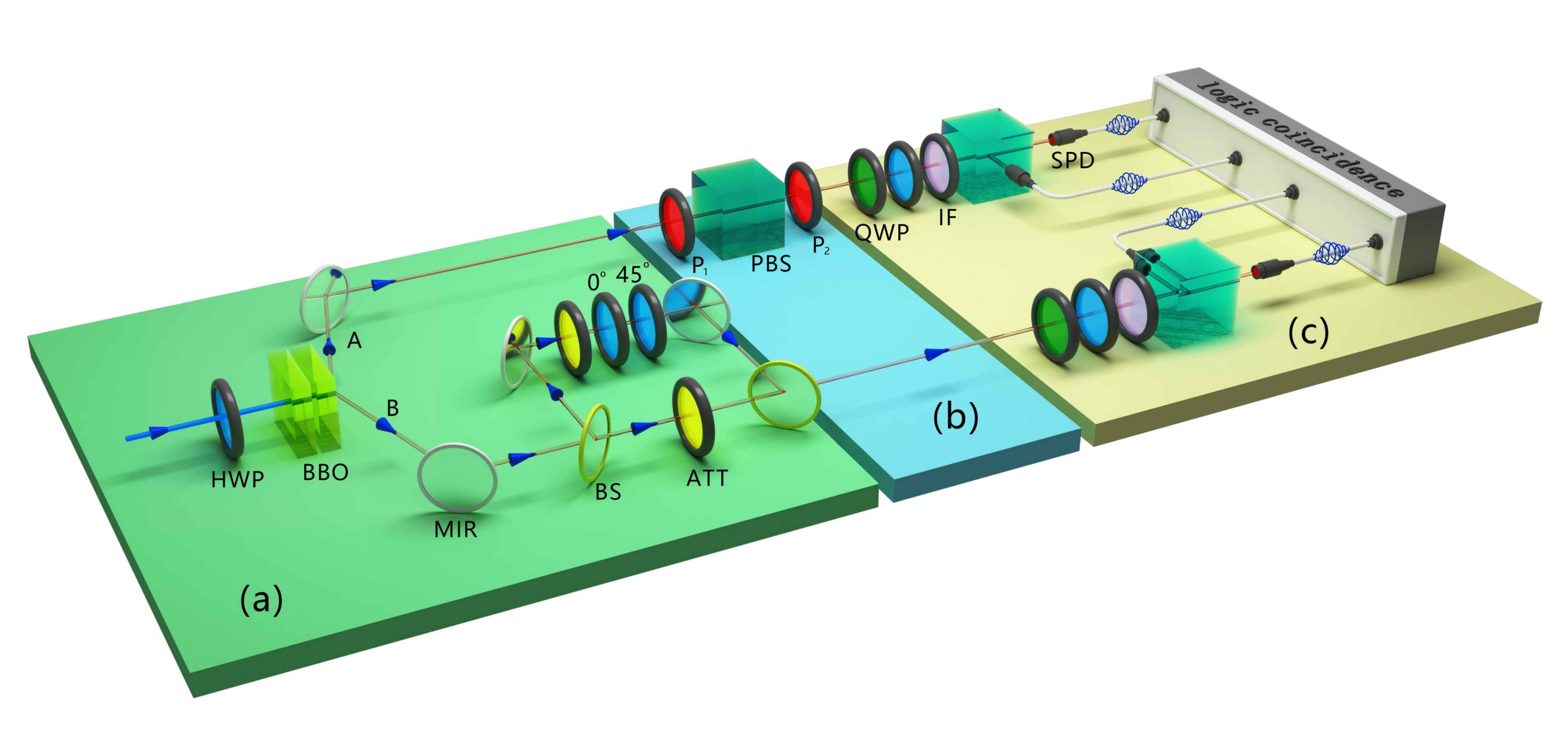}\\
			\caption{Experimental setup. The setup divides into three parts: (a) quantum state preparation, (b) local projective measurement, and (c) quantum state tomography. Part (a) is designed to prepare the initial state as shown in Eq. \eqref{g11} with high fidelity, and to ensure that the state parameters $p$ and  $\theta$ can be easily adjusted. The function of part (b) is to perform local projective measurements on subsystem $A$. The role of part (c) is to reconstruct the initial state or post-measurement state by the tomography process. Component description: HWP, half-wave plate; BBO, -barium borate crystal; MIR, mirror; BS, beam splitter; ATT, attenuator; ${\rm P}_1$ and ${\rm P}_2$, wave plates; PBS, polarizing beam splitter; QWP, quarter-wave plate; IF, interference filter; SPD, single photon detector.}\label{Fig1}
		\end{figure*}
	
	\section{EXPERIMENTAL SCHEME}
		As an illustration, we herein choose a two-qubit mixed state as the initial state, which is with form of
		\begin{equation}\label{g11}
			{{\hat\rho} _{AB}}(p,\theta ) = p\left| {{\Phi _\theta }} \right\rangle \left\langle {{\Phi _\theta }} \right| + (1 - p)\left| {{\Psi _\theta }} \right\rangle \left\langle {{\Psi _\theta }} \right|,
		\end{equation}
		where $\left| {{\Phi _\theta }} \right\rangle  = \cos \theta \left| {HH} \right\rangle  + \sin \theta \left| {VV} \right\rangle $ and $\left| {{\Psi _\theta }} \right\rangle  = \cos \theta \left| {HV} \right\rangle  - \sin \theta \left| {VH} \right\rangle $ are two Bell-like states with the adjustable parameter $\theta  \in [0,{90^ \circ }]$, and
$p \in [0,1]$ is an adjustable mixing parameter which is used to change the proportion of the two Bell-like states. Here, $\left| H \right\rangle $ is encoded the horizontal polarization state and $\left| V \right\rangle $ is encoded the vertical polarization state in a linear optical system \cite{KwiatR773, Xu7, Qi19, Wu454, Li032107}. For the local projective measurement on the subsystem $A$, we resort to the eigenbasis of three Pauli operators $\{ {\sigma _x},{\sigma _y},{\sigma _z}\} $ to
build a most commonly used complete set of MUBs in a qubit system.

		\begin{table}[b]
			\centering
			\caption{The local complete MUB measurements on photon $A$. Here $\{ \left| {{x_i}} \right\rangle ,\left| {{y_i}} \right\rangle ,\left| {{z_i}} \right\rangle \}$ are the eigenbases of the three Pauli operators $\{{\sigma _x}, {\sigma _y}, {\sigma _z}\}$. $\theta_1$ and $\theta_2$ are the optical axis angles of the two wave plates $\rm{P}_1$ and $\rm{P}_2$.}
			\begin{ruledtabular}
				\begin{tabular}{cccccc}
					Operator & Measurement & $\rm{P}_1$ & $\theta_1$ & $\rm{P}_2$ & $\theta_2$\\
					\colrule
					\multirow{2}*{${\sigma _x}$} & $\left| x_0 \right\rangle \left\langle x_0 \right|$ & HWP & $22.5 ^{\circ}$ & HWP & $22.5 ^{\circ}$\\
					~ & $\left| x_1 \right\rangle \left\langle x_1 \right|$ & HWP & $-22.5 ^{\circ}$ & HWP & $-22.5 ^{\circ}$\\
					\hline
					\multirow{2}*{${\sigma _y}$} & $\left| y_0 \right\rangle \left\langle y_0 \right|$ & QWP & $45 ^{\circ}$ & QWP & $-45 ^{\circ}$\\
					~ & $\left| y_1 \right\rangle \left\langle y_1 \right|$ & QWP & $-45 ^{\circ}$ & QWP & $45 ^{\circ}$\\
					\hline
					\multirow{2}*{${\sigma _z}$} & $\left| z_0 \right\rangle \left\langle z_0 \right|$ & HWP & $0 ^{\circ}$ & HWP & $0 ^{\circ}$\\
					~ & $\left| z_1 \right\rangle \left\langle z_1 \right|$ & HWP & $45 ^{\circ}$ & HWP & $45 ^{\circ}$\\
				\end{tabular}
			\end{ruledtabular}		
		\end{table}
		
		We herein set up an all-optical experiment to realize the scheme, and the schematic diagram of our scheme is provided in Fig. \ref{Fig1}.
To be explicit, the experimental setup can be divided into three parts: (a) quantum state preparation, (b) local projective measurement, and (c) quantum state tomography. In part (a), we firstly prepare the entangled mixed state shown as Eq. \eqref{g11}. After a continuous linearly polarized pumped beam with the power of 130 mW and the wavelength of 405nm passes through a half-wave plate (HWP), a beam of polarized light with adjustable horizontal and vertical components $\sin \theta \left| H \right\rangle  + \cos \theta \left| V \right\rangle $ can be harvested. The parameter $\theta $ can be easily adjusted by changing the angle of the optical axis of the HWP. This pumped beam is focused on two type-I $\beta $-barium borate (BBO) crystals ($6.0 \times 6.0 \times 0.5$mm) with optical axis cut at $29.2 ^{\circ}$. After that, a pair of entangled photons with state $\left| {{\Phi _\theta }} \right\rangle  = \cos \theta \left| {HH} \right\rangle  + \sin \theta \left| {VV} \right\rangle $ and the central wavelength of 810nm will be generated by the spontaneous parametric down-conversion (SPDC). In order to prepare the mixed state consisting of two Bell-like state,
we add an unbalanced Mach-Zehnder device (UMZ) in the A-path \cite{Li752}, which includes two 50/50 beam splitters (BSs), two mirrors (MIRs), two attenuators (ATTs), and two HWPs with optical axes $0 ^{\circ}$ and $45 ^{\circ}$, respectively. The mixing weight parameter $p$ can be regulated by the two ATTs. Part (b) is to realize the local projective measurements of A-path photons. It consists of two wave plates $\rm{P}_1$ and $\rm{P}_2$ and a polarizing beam splitter (PBS). By means of different wave plate combinations and setting the corresponding optical axis angles $\theta_1$ (for $\rm{P}_1$) and $\theta_2$ (for $\rm{P}_2$),
the different projective measurements can be implemented. The detailed setting has been provided in Table I \cite{Ding022308}. The role of part (c) is to reconstruct the quantum state by the tomography process, which consists of two quarter-wave plates (QWPs), two HWPs, two 3-nm interference filters (IFs), two PBSs, four single photon detectors (SPDs) and a logic coincidence unit. By setting the optical axis of the QWPs and the HWPs, we can achieve at least 16 groups of measurement bases. Based on the obtained coincidence number, the density matrix of the quantum state can be
 perfectly reconstructed \cite{James052312}.
	
	\section{EXPERIMENTAL RESULTS}
				
		\begin{figure}[t]
			\centering
			\includegraphics[width=8cm]{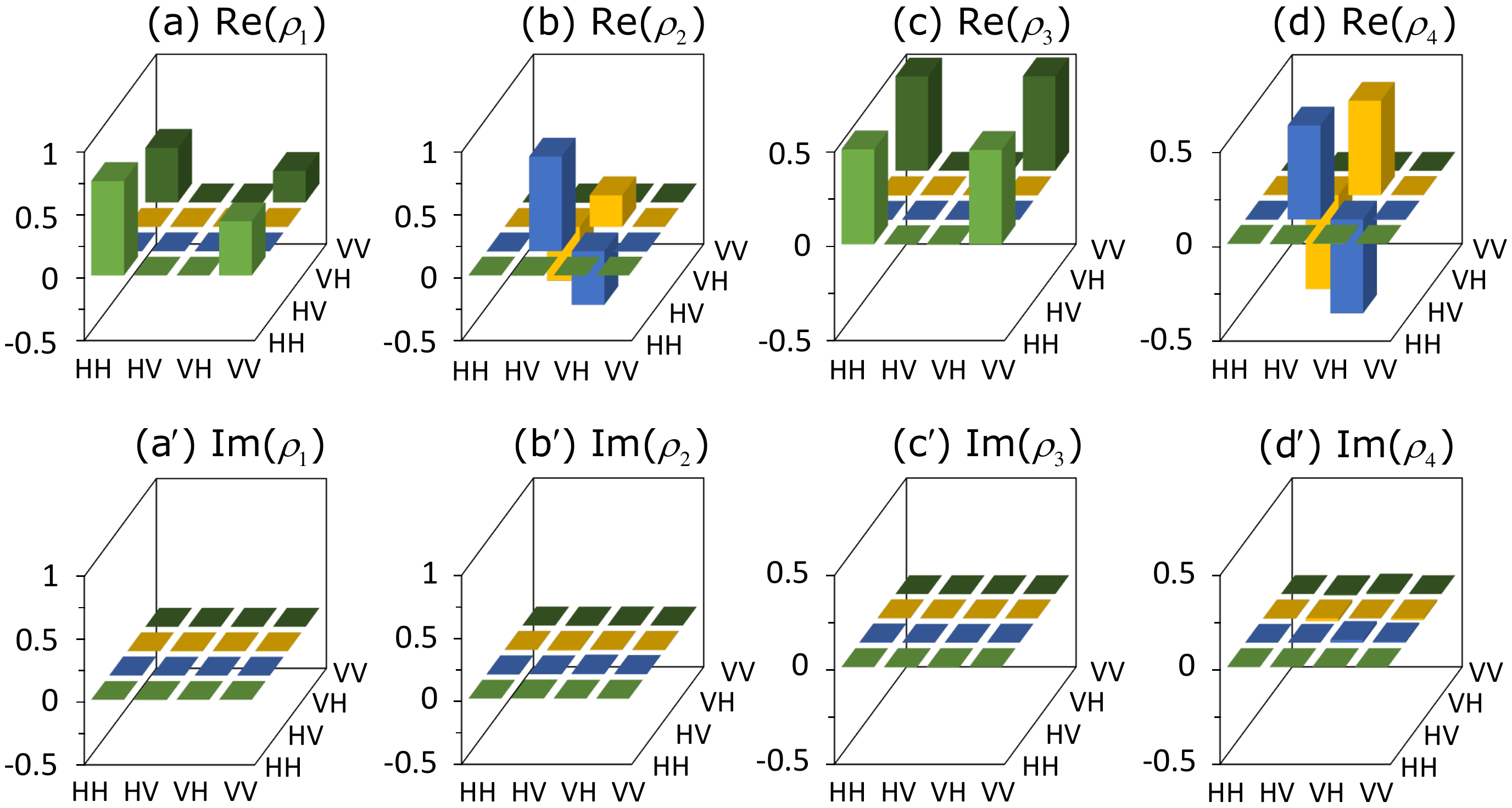}\\
			\caption{Graphical representation of the four Bell-like states. (a) Re(${\hat\rho} _1$), (b) Re(${\hat\rho} _2$), (c) Re(${\hat\rho} _3$), and (d) Re(${\hat\rho} _4$) represent the real parts of the states. (a$'$) Im(${\hat\rho} _1$), (b$'$) Im(${\hat\rho} _2$), (c$'$) Im(${\hat\rho} _3$), and (d$'$) Im(${\hat\rho} _4$) represent their imaginary parts.}\label{Fig2}
		\end{figure}
		
		To prepare the initial states given in Eq. \eqref{g11}, we prepare two kinds of states, i.e., Bell-like states (pure) and Bell-like  diagonal states (mixed).
First, we assure that the parameter $p$ is constant, and choose to prepare two groups of Bell-like states: ${{\hat\rho} _{AB}}(p = 0,\;\theta )$ and ${{\hat\rho} _{AB}}(p = 1,\;\theta )$. In this case, the value of $\theta $ is adopted as $0 ^{\circ}, 10 ^{\circ}, 20 ^{\circ}, 30 ^{\circ}, 40 ^{\circ}, 45 ^{\circ}, 50 ^{\circ}, 60 ^{\circ}, 70 ^{\circ}, 80 ^{\circ}, 90 ^{\circ}$, respectively. Second, we fix the parameter $\theta$ and prepare the other two groups of Bell-like diagonal states: ${{\hat\rho} _{AB}}(p,\;\theta  = 30^\circ )$ and ${{\hat\rho} _{AB}}(p,\;\theta  = 45^\circ )$. Here $p$ is set as $0, 0.1, 0.2, 0.3, 0.4, 0.5, 0.6, 0.7, 0.8, 0.9, 1$, respectively. We reconstruct the density matrices of all initial states by quantum state tomography process shown in part (c) in Fig. \ref{Fig1}. Explicitly, Fig. \ref{Fig2} shows the real and imaginary parts of four typical Bell-like states we prepared: ${{\hat\rho} _1}(p = 1,\;\theta  = 30^\circ ) = \left| {{\Phi _{30^\circ }}} \right\rangle \left\langle {{\Phi _{30^\circ }}} \right|$, ${{\hat\rho} _2}(p = 0,\;\theta  = 30^\circ ) = \left| {{\Psi _{30^\circ }}} \right\rangle \left\langle {{\Psi _{30^\circ }}} \right|$,	${{\hat\rho} _3}(p = 1,\;\theta  = 45^\circ ) = \left| {{\Phi _{45^\circ }}} \right\rangle \left\langle {{\Phi _{45^\circ }}} \right|$,	${{\hat\rho} _4}(p = 0,\;\theta  = 45^\circ ) = \left| {{\Psi _{45^\circ }}} \right\rangle \left\langle {{\Psi _{45^\circ }}} \right|$. By precisely adjusting the mixing proportion of these quantum states, we can easily prepare the Bell-like diagonal states as desired by us. In our scheme, the average fidelity of the experimental result is up to $\bar F({{\hat\rho} _{AB}}) = 0.9983 \pm 0.0025 $, quantified by $ {\rm{Tr}}\sqrt {\sqrt {{{\hat\rho} _{AB}}} {{\hat\rho} _0}\sqrt {{{\hat\rho} _{AB}}} } $, where ${\hat\rho} _0 $ is the corresponding target states, and the error bar is estimated according to the Monte Carlo method \cite{Altepeter105}.
	
		\begin{figure}[t]
			\centering
			\includegraphics[width=8cm]{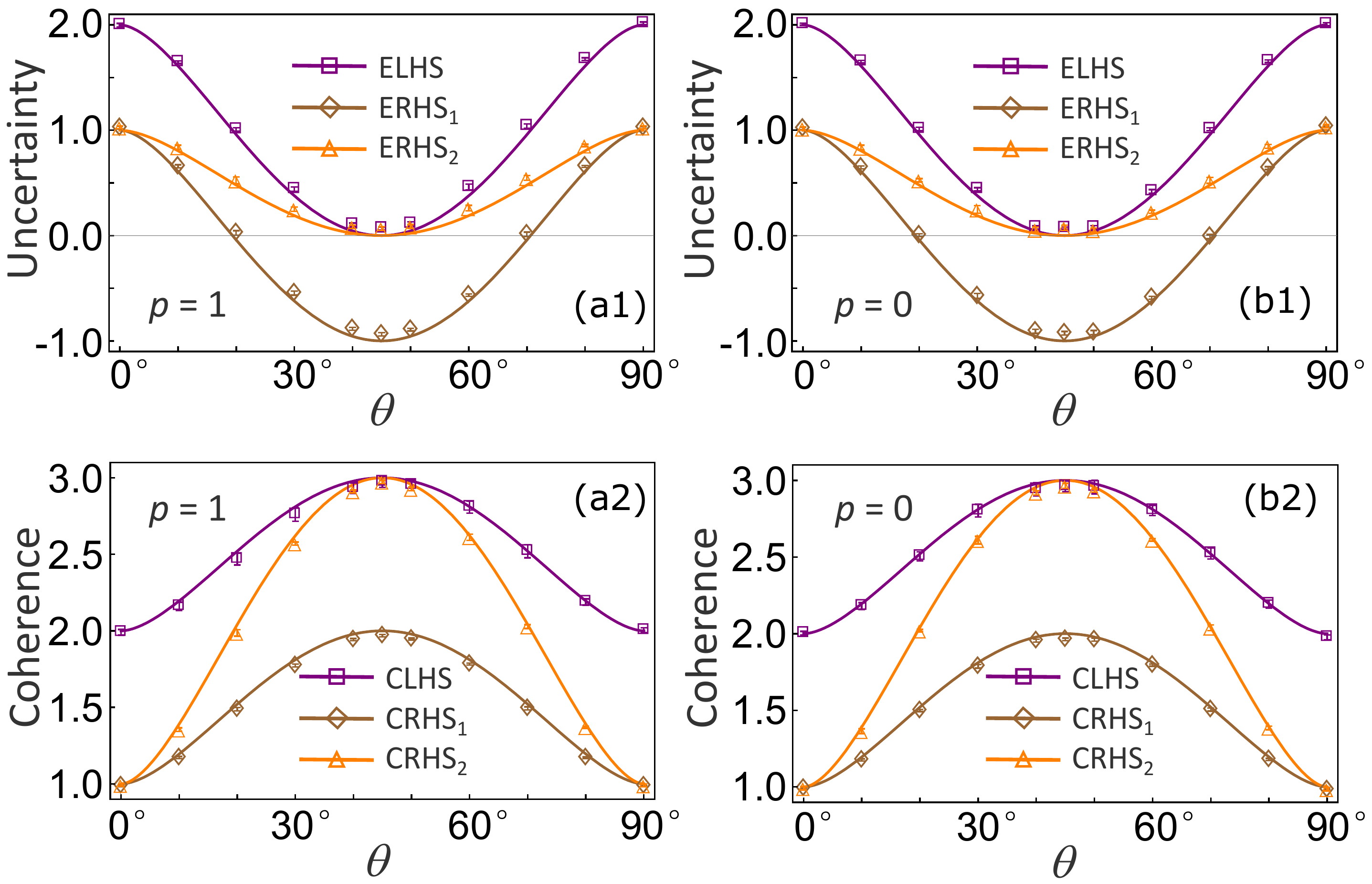}\\
			\caption{Experimental results of the entropic uncertainty relations and the uncertainty relations for quantum coherence under MUBs with the initial Bell-like states: ${{\hat\rho} _{AB}}(p = 0,\;\theta )$ and ${{\hat\rho} _{AB}}(p = 1,\;\theta )$. The $x$ axis denotes the parameter $\theta$ of the initial states. The $y$ axis of (a1) and (b1) denotes the values of the entropic uncertainty, and the $y$ axis of (a2) and (b2) denotes the values of the coherence. The purple squares in (a1) and (b1) denote the measured values of ELHS and the purple squares in (a2) and (b2) denote CLHS. The brown rhombus and the orange triangles in (a1) and (b1) represent the measured values of ERHS1 and ERHS2, while these shapes in (a2) and (b2) represent CRHS1 and CRHS2, respectively. The solid lines represent the corresponding theoretical predictions of uncertainties and coherence, respectively. }\label{Fig3}
		\end{figure}

		With respect to the initial state we prepared, we perform the local complete set of MUB measurements $\{M_x, M_y, M_z\}$ on photon $A$, which are composed of the eigenbases $\{ \left| {{x_i}} \right\rangle ,\left| {{y_i}} \right\rangle ,\left| {{z_i}} \right\rangle \}$ of the three Pauli operators $\{{\sigma _x}, {\sigma _y}, {\sigma _z}\}$: ${M_x} = \{ \left| {{x_i}} \right\rangle \left\langle {{x_i}} \right|\}$, ${M_y} = \{ \left| {{y_i}} \right\rangle \left\langle {{y_i}} \right|\}$ and ${M_z} = \{ \left| {{z_i}} \right\rangle \left\langle {{z_i}} \right|\}$, where $i$ is the outcome of 0 or 1. In linear optics, we generally assume that $\left| {{x_0}} \right\rangle : = \left| D \right\rangle $, $\left| {{x_1}} \right\rangle : = \left| A \right\rangle $, $\left| {{y_0}} \right\rangle : = \left| R \right\rangle $, $\left| {{y_1}} \right\rangle : =\left| L \right\rangle $, $\left| {{z_0}} \right\rangle : = \left| H \right\rangle $, and $\left| {{z_1}} \right\rangle : = \left| V \right\rangle $, where $\left| D \right\rangle  = (\left| H \right\rangle  + \left| V \right\rangle )/\sqrt 2 $, $\left| A \right\rangle  = (\left| H \right\rangle  - \left| V \right\rangle )/\sqrt 2 $, $\left| R \right\rangle  = (\left| H \right\rangle  + i\left| V \right\rangle )/\sqrt 2 $, and $\left| L \right\rangle  = (\left| H \right\rangle  - i\left| V \right\rangle )/\sqrt 2 $ denote the diagonal, antidiagonal, right-circular, and left-circular polarized states \cite{Altepeter105}. After the local measurements, the bipartite states will become ${\hat\rho} _{AB}^{{x_i}}$, ${\hat\rho} _{AB}^{{y_i}}$ and ${\hat\rho} _{AB}^{{z_i}}$ with the corresponding probability ${p_{{x_i}}}$, ${p_{{y_i}}}$ and  ${p_{{z_i}}}$, respectively.
Similarly, we can reconstruct the density matrices of these quantum states according to tomography process,
and work out the corresponding probability via coincidence counts \cite{Li752}. Thereby, one can attain the post-measurement states of the composite system: ${{\hat\rho} _{{M_x}B}} = {p_{{x_0}}}{\hat\rho} _{AB}^{{x_0}} + {p_{{x_1}}}{\hat\rho} _{AB}^{{x_1}}$, ${{\hat\rho} _{{M_y}B}} = {p_{{y_0}}}{\hat\rho} _{AB}^{{y_0}} + {p_{{y_1}}}{\hat\rho} _{AB}^{{y_1}}$ and ${{\hat\rho} _{{M_z}B}} = {p_{{z_0}}}{\hat\rho} _{AB}^{{z_0}} + {p_{{z_1}}}{\hat\rho} _{AB}^{{z_1}}$.	To demonstrate the EURs and the CURs we focus, it is
indispensable to derive the left-hand sides and the right-hand sides of Eqs. \eqref{g08a}, \eqref{g08b}, \eqref{g09}, and
\eqref{g10}, respectively. Explicitly, the left-hand side of Eq. \eqref{g08a} is defined by ${\rm{ELSH}} = S({M_x}|B) + S({M_y}|B) + S({M_z}|B) = S({{\hat\rho} _{{M_x}B}}) + S({{\hat\rho} _{{M_y}B}}) + S({{\hat\rho} _{{M_z}B}}) - 3S({{\hat\rho} _B})$, and the left-hand side of Eq. \eqref{g08b} by ${\rm{CLSH}} = C_{{\rm{re}}}^{{M_x}}({{\hat\rho} _{AB}}) + C_{{\rm{re}}}^{{M_y}}({{\hat\rho} _{AB}}) + C_{{\rm{re}}}^{{M_z}}({{\hat\rho} _{AB}}) = S({{\hat\rho} _{{M_x}B}}) + S({{\hat\rho} _{{M_y}B}}) + S({{\hat\rho} _{{M_z}B}}) - 3S({{\hat\rho} _{AB}})$. ${\rm{ERHS_1}}$ and ${\rm{CRHS_1}}$ stand for the right-hand sides of Eqs. \eqref{g08a} and \eqref{g08b}, ${\rm{ERHS_2}}$ and ${\rm{CRHS_2}}$ represent the right-hand sides of Eqs. \eqref{g09} and \eqref{g10}, respectively.

		\begin{figure}[t]
			\centering
			\includegraphics[width=8cm]{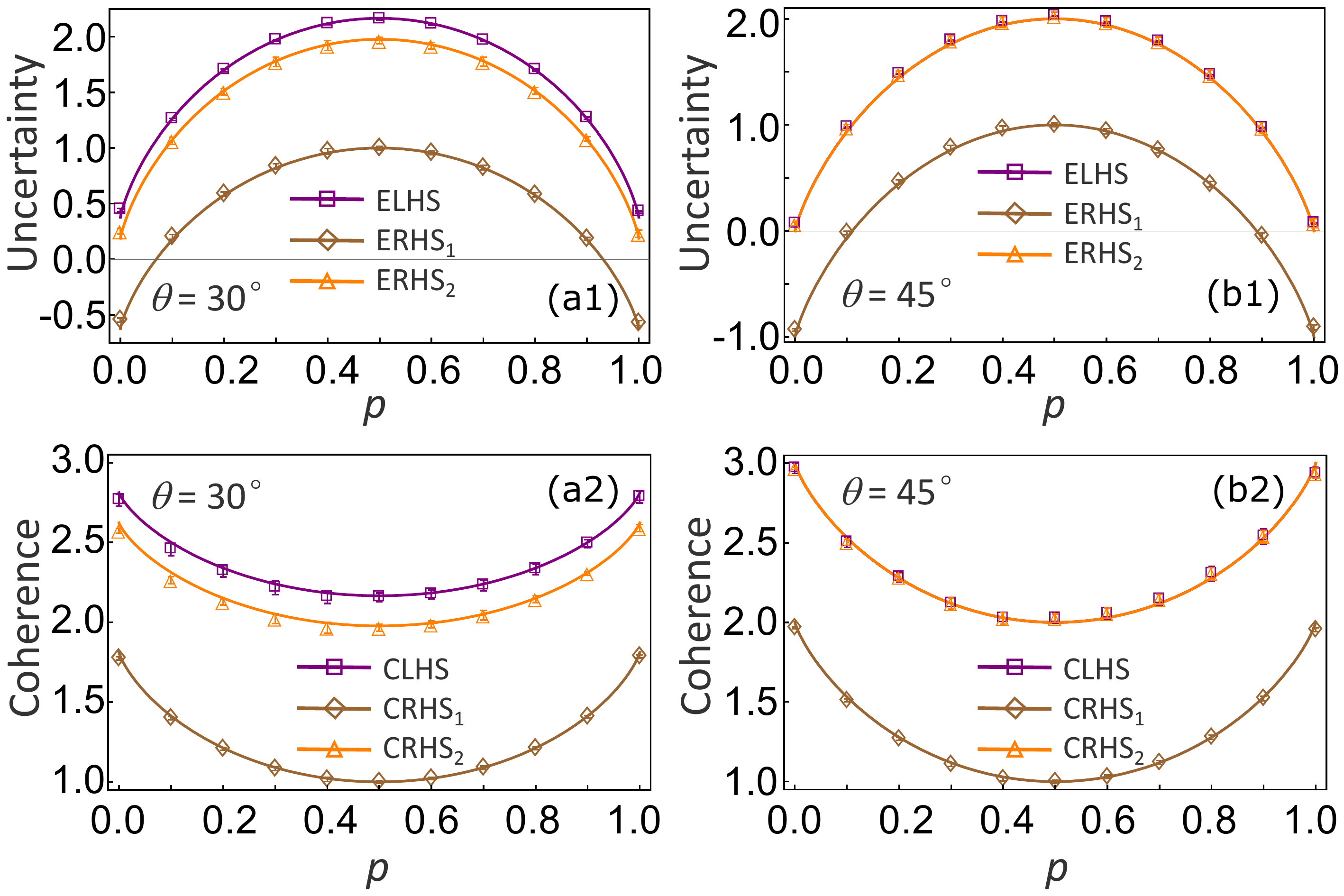}\\
			\caption{Experimental results of the entropic uncertainty relations and the uncertainty relations for quantum coherence under MUBs with the initial Bell-like diagonal states: ${{\hat\rho} _{AB}}(p,\;\theta  = 30^\circ )$ and ${{\hat\rho} _{AB}}(p,\;\theta  = 45^\circ )$. Here, the $x$ axis represents the parameter $p$ of the initial states, and the definitions of the $y$ axis and all symbols   are same as those in FIG. \ref{Fig3}.}\label{Fig4}
		\end{figure}

		The experimental results and theoretical predictions of the EURs and the CURs in MUBs with the initial Bell-like states and Bell-like diagonal states have been shown in Figs. \ref{Fig3} and \ref{Fig4}. The $x$ axis in Fig. \ref{Fig3} represents the parameter $\theta$ of the initial states, while the $x$ axis in Fig. \ref{Fig4} represents the parameter $p$.
In both figures, the $y$ axis of subgraphs (a1) and (b1) represents the magnitude of the uncertainty, and the $y$ axis of subgraphs (a2) and (b2) represents the values of the coherence. The purple squares in (a1) and (b1) represent the measured values of ELHS and the purple squares in (a2) and (b2) represent CLHS. The brown rhombus and the orange triangles in (a1) and (b1) denote the measured values of $\rm{ERHS}_1$ and $\rm{ERHS}_2$, while these shapes in (a2) and (b2) denote $\rm{CRHS}_1$ and $\rm{CRHS}_2$, respectively. The solid lines with different colors represent the corresponding theoretical predictions of uncertainties and coherences, respectively. Following the  figures, it has been directly shown that  the experimental results coincide with the theoretical predictions very well, which lies in:
		
		(i) All experimental results are consistent with the theoretical curve within the error range, which not only verifies the theory in an all-optical setup, but also indicates that the prepared quantum states have high fidelity.
		
		(ii) The orange triangles ($\rm{ERHS}_2$ or $\rm{CRHS}_2$) are always above the brown rhombus ($\rm{ERHS}_1$ or $\rm{CRHS}_1$), which means that the lower bounds of the entropic uncertainty relations and the coherence uncertainty relations  can be improved by the Holevo quantity and mutual information. In particular, for Bell diagonal states ${{\hat\rho} _{AB}}(p,\;\theta  = 45^\circ )$ in Figs. \ref{Fig4} (b1) and (b2), it indicates that the orange rhombus and the purple squares (ELHS or CLHS) almost coincide, meaning that the lower bounds of the EURS and CURs are enhanced.
		
		(iii) With the variation of quantum state parameters, the total entropic uncertainty (ELHS) will increases (decreases), while the total coherence uncertainty (CLHS) decreases (increases). It means that the entropic uncertainty is inversely correlated with the coherence, which is essentially in agreement with the conclusion from \cite{Dolatkhah13}. For the Bell-like states ${{\hat\rho} _{AB}}(p = 0,\;\theta )$ and ${{\hat\rho} _{AB}}(p = 1,\;\theta )$ in Figs. \ref{Fig3},	the total coherence reaches a maximum of 3 at $\theta  = 45^\circ$, while the uncertainty decreases to the minimum value of 0. This phenomenon is also shown in Figs. \ref{Fig4} (b1) and (b2) with Bell diagonal states ${{\hat\rho} _{AB}}(p,\;\theta  = 45^\circ )$.

	\section{CONCLUSIONS}
	
		To conclude, we experimentally demonstrated the entropic uncertainty relations and the coherence-based uncertainty relations via an all-optical platform. We
prepare the initial states in Bell-like states and Bell-like diagonal states with high   fidelity $(\sim 99.83\%)$. We perform a complete set of MUBs measurements on one of the subsystems, i.e., three general Pauli-operator measurements. By virtue of quantum tomography, we reconstruct the density matrices of the initial states and the post-measurement states,
as well as gaining the corresponding measurement probability. Therefore, we can easily attain the magnitude of the uncertainty and the lower bounds in those proposed uncertainty inequalities both experimentally and theoretically. Remarkably, our experimental results coincide with the theoretical predictions very well. Moreover, it also verifies that the lower bounds of these inequalities are effectively improved by means of the Holevo quantity and mutual information. Further, the experimental results also suppose that the entropic uncertainty is inversely correlated with the quantum coherence.	
With these in mind, we believe that our demonstration could deepen the understanding of entropic uncertainty relations and its connection to quantum coherence,
and the results are expected to be applicable to quantum key distributions.

	\section*{ACKNOWLEDGMENTS}
		This work was supported by the National Natural Science Foundation of China (Grant Nos. 11575001, 11405171, 61601002 and 11605028), the Natural Science Research Project of Education Department of Anhui Province of China (Grant No. KJ2018A0343), the Key Program of Excellent Youth Talent Project of the Education Department of Anhui Province of China (Grant Nos. gxyqZD2019042, gxyq2018059 and gxyqZD2018065), the Key Program of Excellent Youth Talent Project of Fuyang Normal University (Grant No. rcxm201804), the Open Foundation for CAS Key Laboratory of Quantum Information (Grant Nos. KQI201801 and KQI201701), and the Research Center for Quantum Information Technology of Fuyang Normal University (Grant No. kytd201706).
				
		Zhi-Yong Ding and Huan Yang contributed equally to this work.
	
	\bibliographystyle{plain}
	
\end{document}